\documentclass[prl,twocolumn,superscriptaddress,showpacs]{revtex4}
\usepackage{graphicx}
\usepackage{amssymb}
\begin{document}

\title{Collective excitations of trapped one-dimensional dipolar quantum gases}

\author{P. Pedri}
\affiliation{Laboratoire de Physique Th\'eorique e Mod\`eles Statistiques, Universit\'e Paris-Sud, Orsay, France}
\author{S. De Palo}
\affiliation{DEMOCRITOS INFM-CNR and Dipartimento di Fisica Teorica, Universit\`a Trieste, Trieste, Italy }
\author{E. Orignac}
\affiliation{Laboratoire de Physique de l'\'Ecole Normale
  Sup\'erieure de Lyon, CNRS-UMR5672, Lyon, France}
\author{R. Citro}
\affiliation{Dipartimento di Fisica "E. R. Caianiello" and CNISM, Universit\`a degli Studi di Salerno, Salerno, Italy}
\author{M.~L. Chiofalo}
\affiliation{Classe di Scienze, INFN and CNISM, Scuola Normale
  Superiore, Pisa, Italy}
\affiliation{Centre \'Emile Borel, Institut Henri Poincar\'e, Paris, France}

\begin{abstract}
We calculate the excitation modes of a 1D dipolar quantum gas confined
in a harmonic trap with frequency $\omega_0$ and 
predict how the frequency of the breathing $n=2$ mode characterizes
the interaction strength evolving from the Tonks-Girardeau value
$\omega_2=2\omega_0$ to the quasi-ordered, super-strongly interacting
value $\omega_2=\sqrt{5}\omega_0$. Our predictions are obtained
within a hydrodynamic Luttinger-Liquid theory after 
applying the Local Density Approximation to the
equation of state for the homogeneous dipolar gas, which are in turn determined from
Reptation Quantum Monte Carlo simulations. They are shown to be in quite accurate 
agreement with the results of a sum-rule approach.  
These effects can be observed in current experiments, revealing the Luttinger-liquid 
nature of 1D dipolar Bose gases. 
\end{abstract}
\maketitle

\smallskip

\paragraph{Introduction.-}
The bottom line in the design of conceptually novel technological
applications is the possibility of reaching extreme quantum
degeneracy under controlled conditions. This is quite a remarkable
property of ultracold atomic gases, which is being evidenced by
the interdisciplinary contribution of quantum optics, quantum
information, condensed matter, atomic, and fundamental physics.
Extreme quantum limit can here be obtained after lowering the
temperature down to the nanokelvin scale, and by tuning the atomic
interactions and the dimensionality. Dimensionality can indeed be
reduced to two and one dimensions (1D) after using a variety of
magnetic and optical techniques including optical
lattices~\cite{dimensionality_1,dimensionality_2,Bloch_1D,dimensionality_3}.
The interactions can also be manipulated almost at will in their
short-range part by means of the Fano-Feshbach resonance
mechanism~\cite{ff_1,ff_2,ff_exp_1,ff_exp_2,ff_exp_3}. In the case
of atomic or molecular species with large magnetic or electric
moments, proposals have been put forward which predict the
possibility of tuning the long-range dipolar tail of the
interaction~\cite{giovanazzi_pfau,zoller_buchler}. The observation
of dipolar interactions~\cite{cr_dipolar} in atomic $^{52}$Cr
vapors with relatively large magnetic moment $\mu\simeq 6\mu_B$,
is especially promising for applications, since it opens the way
to the realization of {\it e.g.} different quantum phases
~\cite{applications_1,applications_2} and the observation of
spin-charge separation~\cite{kollath}. Molecular dipolar crystals
have been more recently proposed as a realization of high fidelity
quantum memory for quantum computation~\cite{rabl_zoller}. In a
set of landmark experiments
~\cite{dipolar_theo_earlier_1,dipolar_theo_earlier_2} and
supported by simulational
work~\cite{dipolar_theo_recent_1,dipolar_theo_recent_2}, vapors of
$^{52}$Cr atoms have been Bose-condensed~\cite{cr_bec}. 
Yet, the effect of the dipolar interaction has been largely
enhanced after reducing the strength of the short-range
part~\cite{pfau_a0}.

Combination of 1D geometries with tunable interactions may provide
an easy access to enhanced quantum correlations. Experimental
realizations of the strongly correlated Tonks-Girardeau
(TG) gas~\cite{TG} have already been observed in atomic vapors
with contact interactions~\cite{Bloch,Weiss}. Going beyond the TG
regime under these conditions is not an easy task~\cite{giorgini}. 
However, as we have more recently demonstrated by
a combined Reptation Quantum Monte Carlo (RQMC) and bosonization
approach~\cite{dipolar_citro_pra_rc}, a homogeneous 1D
Bose gas with dipolar interactions 
is a very strongly correlated Luttinger liquid
with the parameter $K<1$~\cite{giamarchi_general_ref} at all
densities $nr_0\gtrsim 0.1$ with $r_0$ being the range of the
dipolar potential (see below). The Luttinger liquid crosses over
from a Tonks-Girardeau gas setting in for $nr_0\lesssim 0.1$ to a
high-density quasi-ordered state~\cite{astra} which can be viewed
as the analogue of a Charge Density Wave. We have also discussed
how the use of polar molecules may provide access to this
quasi-ordered state, which from now on we refer to as Dipolar
Density Wave (DDW).

The knowledge of the interaction regime
is a basic tool for all conceivable applications.
Since ongoing experiments will be performed in confined geometry,
one of the best suited and controlled methods is to exploit the collective
excitations related to the discrete modes in the harmonic
trap~\cite{theo_string}, which has indeed been largely used since the very
first experiments~\cite{collective_exc_exp_1,collective_exc_exp_2}.
The study of the collective modes can also be useful in view of the
proposal to implement the quantum memory, to reveal and investigate
decoherence mechanisms possibly arising from
the coupling of internal and external degrees of freedom on the
DDW state~\cite{rabl_zoller}.

In this Letter we predict
for the first time the
crossover behavior of the collective modes of the trapped 1D dipolar Bose gas
by two different theoretical methods, namely a sum-rule approach
and a hydrodynamic Luttinger-liquid model. The two methods
share the application of the Local Density Approximation to
the equation of state as determined from our RQMC simulational 
data, and are shown
to give results in quite remarkable agreement.

\paragraph{The equation of state.-}
\noindent We first
determine the ground-state energy per particle 
(within the statistical error)
of the homegeneous dipolar Bose gas by resorting to
Reptation Quantum Monte Carlo simulations (RQMC)~\cite{RQMC}.
As described in more detail
in Ref.~\cite{dipolar_citro_pra_rc}, we consider $N$ atoms or molecules of
mass $M$ and permanent dipole moments
arranged along a line in the limit
of negligible contact interaction~\cite{pfau_a0} 
and polarized in the orthogonal direction.
The Hamiltonian $H=\left(-(nr_0)^2
\sum ({\partial^2}/{\partial x^2})+
(nr_0)^3 \sum_{i<j}|x_i-x_j|^3 \right)$
is defined in effective Rydberg
units $Ry^*={\hbar^2}/({2 M r_0^2})$ with
$r_0\equiv M C_{dd}/(2
\pi \hbar^2)$ the effective Bohr radius, 
$C_{dd}=\mu_0\mu_d^2$
or $C_{dd}=d^2/\epsilon_0$ the interaction stengths 
for magnetic $\mu_d$ or electric $d$ dipole moments
respectively. 
The governing dimensionless parameter is $n  r_0$, with 
$n$ the number of dipoles per unit length. 

The RQMC data for the dependence of the energy per particle 
$\varepsilon$ on
$n$  are  fitted by the expression
$\varepsilon(nr_0)/Ry^*=[\zeta(3)(nr_0)^4+a(nr_0)^e+b(nr_0)^f+c(nr_0)^{2+g}]
[1+nr_0]^{-1}+(\pi^2/3)(nr_0)^2
[1+d(nr_0)^g]^{-1}$,
where $a=3.1(1)$, $b=3.2(2)$, $c=4.3(4)$, $d=1.7(1)$,
$e=3.503(4)$, $f=3.05(5)$, and $g=0.34(4)$ with an overall
$\chi_{red}^2\simeq 5$. For $nr_0 \ll 1$, 
$\varepsilon(nr_0)/Ry^*\sim (\pi^2/3)(nr_0)^2$ and for $n r_0 \gg 1$, 
$\varepsilon(nr_0)/Ry^* \sim \zeta(3)(nr_0)^3$ so that both the TG and
DDW limits are satisfied. The functional form 
for $\varepsilon(n)$ here provided can be used in further 
calculations. Here, we use it to obtain the  chemical
potential $\mu_{\text{RQMC}}(n)=(1+n(\partial/\partial
n))\varepsilon(n)$.

\paragraph{The Local Density Approximation.-} \noindent
In experiments, the
effectively 1D dipolar gas is confined by a harmonic potential
$V(x)=M\omega_0^2x^2/2$ in the axial direction. The dynamical
behavior of the confined gas can be found from hydrodynamic equations
at temperature $T=0$ using a Local Density Approximation (LDA) to the
equation of state, in which the energy of the inhomogeneous
system is a local functional of the local density
$n(x)$ expressed as the integral over $x$ 
of the energy density of the homogeneous
gas with density $n(x)$. 
The validity of such an approach is limited to dynamical
effects in which the typical length over which $n(x)$ varies is 
much larger than  the average
interparticle distance in the axial direction. Furthermore, the dynamical
behavior in the radial (transverse) plane must remain frozen. 
Under these
conditions, the ground-state density profile $n(x)$
of the trapped dipolar gas is obtained by 
plugging $\mu_{\text{RQMC}}(n)$ into the equation
\begin{equation}
\mu(n(x))+V(x)=\mu_0
\label{eq:tf_lda}\; ,
\end{equation}
and solving it
for $\mu_0$ and $n(x)$ with the condition 
that the total number of particles $N=\int_{-R}^R n(x) dx$ is conserved, 
where the Thomas-Fermi radius $R=(2\mu_0/(M\omega_0^2))^{1/2}$
is such that $n(\pm R)=0$. The condition can be casted in the form
$N(r_0/a_{ho})^2=
(\mu_0/Ry^*)^{1/2}\int_{-1}^1 r_0\mu^{-1}[(\mu_0/Ry^*)(1-x^2)]dx$,
where
$a_{ho}=(\hbar/M\omega_0)^{1/2}$ is the harmonic oscillator length.
This identifies $N(r_0/a_{ho})^2$ as the interaction parameter
driving the trapped dipolar gas from the
TG across the DDW regime.

The evolution of the
calculated density profiles 
through the crossover, shows the expected
increase of the central density $n(0)$ with  the chemical potential
$\mu_0$, as well as a steepening of the profiles  at the trap edges. 
The calculated  profiles agree with the analytical results expected
in the TG and in the DDW limits.

\paragraph{The breathing mode from a sum rule (SR) approach.-}
\noindent We
first determine the effect of the interactions on the frequency of the lowest
compressional (breathing) mode of the trapped gas by a sum-rules
approach. This mode is coupled to the ground state by the operator
$\hat{X}_2=\sum_{i=1}^N x_i^2$.  
This makes it the easiest to probe in current experiments, as it is
excited by modulating the trap frequency and observed
by following the time-evolution of the width of the cloud
by conventional density-imaging techniques. 
The frequency $\omega_B$
of the breathing mode satisfies an inequality that can be derived from
the sum
rule~\cite{Stringa_Pita_book,Chiofalo_sr_charged_bose,menotti_stringari}
$\omega_B^2\le m_1/m_{-1}$, where $2m_1=\langle[X,[H,X]]\rangle$ and
$2m_{-1}$ is the static response function. After some simplification,
this yields the upper bound:  
\begin{equation}
 \omega_B^2 \le \Omega_B^2= -2\frac{\langle x^2 \rangle}{\partial \langle x^2\rangle/\partial \omega_0^2}\;,
\label{eq:wb_sr}
\end{equation}
where $\langle x^2 \rangle=N^{-1}\int x^2 n(x)dx$. A closed form
expression of $\Omega_B$ in Eq.~(\ref{eq:wb_sr}) can be obtained by means of
a scaling argument whenever $\mu(n)$ is of
the form $\mu(n)=\lambda n^\gamma$, resulting in $\Omega_B=
\omega_0 (2+\gamma)^{1/2}$. In particular, in the TG case with
$\gamma=2$
this yields $\Omega_B = 2\omega_0$, and in the DDW case with
$\gamma=3$ it yields $\Omega_B\leq \sqrt{5} \omega_0$.  

\begin{figure}
\includegraphics[width=70mm,angle=-90]{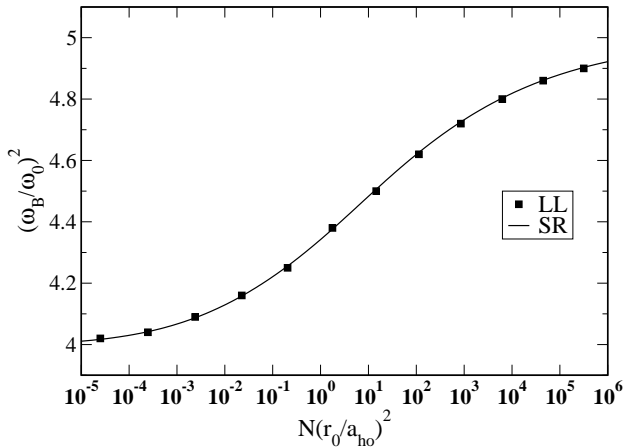}
\caption{Squared frequency $\omega_B^2$ of the breathing mode scaled to the
trap frequency $\omega_0$ {\it vs.} the interaction parameter
$N(r_0/a_{\rm ho})^2$, as calculated from two models. 
Solid line: sum-rule approach Eq.~(\ref{eq:wb_sr}) (SR). 
Symbols: Luttinger-liquid 
hydrodynamics Eq.~(\ref{eq:eigen}) (LL).}
\label{fig:wb}
\end{figure}

For intermediate interaction strengths,
we have to resort to a numerical estimation of
Eq.~(\ref{eq:wb_sr}) using the LDA density profile. The result is
represented by the solid line in Fig.~\ref{fig:wb}, 
showing the smooth evolution
of the  breathing mode frequency $\omega_B$ from the TG to the DDW regimes.
Figure~\ref{fig:wb} represents an useful tool to identify the
interaction regime of the trapped dipolar gas through one of
the best handled experimental probes available with cold
atomic (molecular) quantum
gases~\cite{collective_exc_exp_1,collective_exc_exp_2}. However, 
Eq.~(\ref{eq:wb_sr}) in principle yields an upper bound for the
frequency of the lowest compressional mode. 
Moreover, estimating the frequencies of the higher modes 
by the sum-rule approach is not as simple. 
We thus switch to an alternative, hydrodynamic, approach to compute 
their frequencies.  

\paragraph{The excitation modes from hydrodynamic Luttinger equations.-}
 In~\cite{dipolar_citro_pra_rc}, we have shown
that the low-energy behavior
of a homogeneous dipolar Bose gas is well described by the Luttinger
hamiltonian, and we have determined the density-dependence of the
velocity $u$ and Luttinger exponent $K$
by combining bosonization and RQMC techniques. The $u$ and $K$ obtained from 
the RQMC structure factor were found to agree with those extracted 
from the RQMC energy {\it via} the Luttinger relations
$uK=M^{-1}\pi n$ and $u/K=\pi^{-1}\partial_n\mu(n)$ 
embodying Galilean invariance~\cite{haldane_bosons,giamarchi_book_1d}. 
We now assume that in a
slowly varying  external trapping potential, the dipolar Bose gas
can be described by a Luttinger liquid hamiltonian,
\begin{eqnarray}\label{eq:hamiltonian}
  H_{LL}=\int_{-R}^R \frac{dx}{2\pi} \left[ u(x) K(x) (\pi\Pi)^2 +
  \frac{u(x)}{K(x)} (\partial_x \phi)^2\right],
\label{eq:LL_inho}
\end{eqnarray}
where  $u(x)$ and $K(x)$ now depend on position {\it via} the LDA
$n(x)$. They are related through
$u(x) K(x)=M^{-1}\pi n(x)$ and $u(x)/K(x)=\pi^{-1}\partial_n
\mu(n)|_{n=n(x)}$, extending the Luttinger-liquid relations 
to the weakly inhomogeneous system. In Eq.~(\ref{eq:LL_inho}),
$\phi$ and $\Pi$ satisfy canonical commutation relations
$[\phi(x),\Pi(y)]=i\delta(x-y)$, leading to the equations of
motion:
\begin{eqnarray}
\label{eq:eom1}
\partial_t \phi(x,t) &=& \pi u(x) K(x)  \Pi(x,t), \\
\label{eq:eom2}
\pi \partial_t \Pi(x,t) &=& \partial_x \left(\frac{u(x)}{K(x)}
\partial_x \phi(x,t) \right)\; .
\end{eqnarray}
These equations of motion must be complemented by boundary
conditions expressing that no current flows across the edges, {\it i. e.}
$j(\pm R,t)=0$. Since $\delta n=-\partial_x\phi/\pi$, the continuity
equation $\partial_t (\delta n) + \partial_x j=0$ leads naturally  to
$j=\partial_t \phi/\pi$, and allows us to rewrite the
boundary conditions as $\phi(-R)=\phi_0$ and
$\phi(R)=\phi_1$~\cite{kecke05_trapped}.

Eqs.~(\ref{eq:eom1})--(\ref{eq:eom2}) possess a stationary solution,
$\partial_x \phi\propto K(x)/u(x)$. This solution actually
describes the addition of one particle to the system  $\phi(x) \to
\phi(x)-\pi g(x)/g(R)$ where $g(x)=\int_{-R}^x (K(y)/u(y))dy$. A
consequence is that $\phi_0$ and $\phi_1$ are not independent, but
related through $\phi_1-\phi_0=-\pi N$ with $N$ being the number of
particles added to the system. This form generalizes the
bosonization formula derived in the case of a homogeneous
system~\cite{haldane_bosons}. Alternatively, the static solution can be
derived by considering the effect of a perturbation to the
trapping potential~\cite{ed_long}. Combining
Eqs.~(\ref{eq:eom1})--(\ref{eq:eom2}) and using linearity to search
for solutions of the form $\phi(x,t)=\phi_0 -\pi N g(x)/g(R)
+\sum_{n\ge 1} \varphi_n(x)e^{i\omega_n t}$ we find:    
\begin{eqnarray} \label{eq:eigen}
  -M \omega_n^2 \varphi_n = n(x) \partial_x \left(\partial_n\mu(n(x))
    \partial_x \varphi_n \right).
\end{eqnarray}
with the boundary conditions $\varphi_n (\pm R)=0$ for the
discrete Fourier components $\varphi_n$. 

While Eq.~(\ref{eq:eigen}) is cast in a form
identical to the 
hydrodynamic equation for density excitations~\cite{menotti_stringari}, 
we remark that here it has been 
derived from $H_{LL}$ Eq.~(\ref{eq:LL_inho}). 
Thus, a comparison of the measured excitation frequencies with those
predicted from (\ref{eq:eigen}) with our RQMC data,  
provides a test of Luttinger-liquid behavior in trapped 1D dipolar Bose gases 
within the validity of LDA.    

In the case of a harmonic trapping, one of the eigenfrequencies in
Eq.~(\ref{eq:eigen}) is obtained straightfowardly. Indeed,
substituting $\varphi_1(x) \propto n(x)$ in Eq.~(\ref{eq:eigen}),
differentiating Eq.~(\ref{eq:tf_lda}) with respect to $x$ and using
the Luttinger-liquid relations, we find that $\varphi_1(x)$ is an
eigenfunction of~(\ref{eq:eigen}) associated with the eigenvalue
$\omega_0^2$. This particular solution is simply the Kohn mode or
sloshing mode describing a center-of-mass oscillation. Indeed, it can
be recovered by expanding to first order in $A$ the expression
$\phi(x,t)=\phi(x-A\cos(\omega_0 t),0)$ which describes a center of
mass motion with rigid density.
The eigenfrequencies are exactly known also in the two asymptotic limits. 
Indeed, insertion of $\mu(n)\propto n^\gamma$ 
in Eqs.~(\ref{eq:tf_lda}) and~(\ref{eq:eigen}), yields 
solutions  of the form $\varphi_n(x)=A_n(1-x^2/R^2)^{1/\gamma}
C_n^{(1/\gamma+1/2)}(x/R)$, the associated 
eigenvalues being  $\omega_n^2=\omega_0^2
n[2+(n-1)\gamma]/2$
~\cite{menotti_stringari,ed_long}, where $C_n^{(\alpha)}$ are Gegenbauer
polynomials and $A_n$ normalization factors~\cite{petrov04_bec_review},  
Thus, at the two
opposite TG ($\gamma=2$) and DDW ($\gamma=3)$ limits one finds
respectively $\omega_n=n^2 \omega_0^2$ and
$\omega_n=n(3n-1)\omega_0^2/2$. 
For intermediate densities, 
we have solved Eq.~(\ref{eq:eigen}) using the {\it Sledge}
algorithm available online~\cite{sledge}, after inserting as
ingredients the computed LDA density profiles from 
Eq.~(\ref{eq:tf_lda}) and of the analytical expression
for $\partial_n\mu$ obtained from the RQMC fit, evaluated at the
local density for different values of the interaction parameter
$N(r_0/a_{ho})^2$. In the numerical solution we have taken special 
care of the finite mesh-size effects for best accuracy. 

\begin{figure}[tb]
\includegraphics[width=70mm,angle=-90]{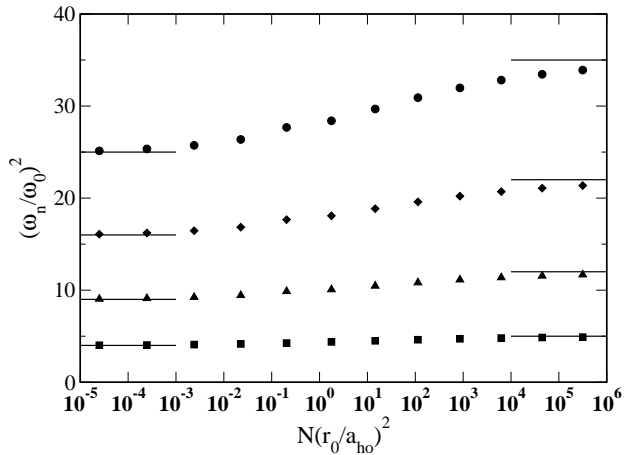}
\caption{$(\omega_n/\omega_0)^2$ 
{\it vs.} $N(r_0/a_{aho})^2$ from Eq.~(\ref{eq:eigen}).
From bottom to top: the modes with $n=2$, $3$, $4$, and $5$. Tick solid lines:
limiting values in the TG and DDW regimes at each $n$.}
\label{fig:modes}
\end{figure}

The values of the breathing frequency $\omega_B/\omega_0$ obtained
from Eq.~(\ref{eq:eigen}) are represented in
Fig.~\ref{fig:wb} by the filled symbols, and agree up to 
the second digit with the sum-rule result. This was expected 
in the TG and DDW regimes where,  
as already noticed in
Ref.~\cite{menotti_stringari}, an equality sign holds in
Eq.~(\ref{eq:wb_sr}). 
The eigenfrequencies of the
higher modes with $n=3$, $4$, and $5$ are displayed in
Fig.~\ref{fig:modes} as functions of $N(r_0/a_{ho})^2$, showing
the same smooth crossover behavior between the two opposite TG and
DDW regimes. The exact frequencies in these asymptotic regimes are
recovered by the numerical calculations. 

\paragraph{Conclusions.-} In conclusion, we have predicted the
evolution of the collective modes of a 1D dipolar Bose gas through 
the crossover
from the Tonks-Girardeau to the Dipolar-Density-Wave regime. These modes, and
especially the breathing mode can be excited and measured by quite standard
and reliable techniques. Our results are relevant to experiments,
where they can be used to determine the interaction
regime, and extended to investigate the occurrence of
decoherence effects in quantum applications~\cite{rabl_zoller}. 
From the theoretical point of view, our results confirm for the trapped gas 
the smooth crossover from the TG to the DDW regimes
~\cite{dipolar_citro_pra_rc}, as expected for 
weak inhomogeneity. The RQMC functional form 
of the energy per particle that we explicitly provide in this work, 
can be useful to investigate the issue, crucial for applications, 
of the crystal phase stabilization by means of 
{\it e.g.} a commensurate, though shallow, 
optical lattice. 

MLC would like to thank the Institut Henri Poincar\'e - Centre Emile 
Borel in Paris for hospitality and support. This work was suported by the 
Minist\`ere de la Recherche (grant ACI Nanoscience 201),
by the ANR (grants NT05-42103 and 05-Nano-008-02) and by the IFRAF Institute.


\begin{thebibliography}{10}

\bibitem{dimensionality_1}
D.~{Hellweg} et~al.,
\newblock Applied Physics B {\bf 73}, 781 (2001).

\bibitem{dimensionality_2}
A.~{G{\"o}rlitz} et~al.,
\newblock Phys. Rev. Lett. {\bf 87}, 130402 (2001).

\bibitem{Bloch_1D}
I.~Bloch,
\newblock Nature Phys. {\bf 1}, 23 (2005),
\newblock and refs. therein.

\bibitem{dimensionality_3}
S.~{Richard} et~al.,
\newblock Phys. Rev. Lett. {\bf 91}, 010405 (2003).

\bibitem{ff_1}
H.~Feshbach,
\newblock Ann. Phys. {\bf 5}, 357 (1958).

\bibitem{ff_2}
U.~Fano,
\newblock Phys. Rev. {\bf 124}, 1866 (1961).

\bibitem{ff_exp_1}
S.~{Inouye} et~al.,
\newblock Nature {\bf 392}, 151 (1998).

\bibitem{ff_exp_2}
J.~L. {Roberts} et~al.,
\newblock Phys. Rev. Lett. {\bf 86}, 4211 (2001).

\bibitem{ff_exp_3}
T.~{Weber}, J.~{Herbig}, M.~{Mark}, H.-C. {N{\"a}gerl}, and R.~{Grimm},
\newblock Science {\bf 299}, 232 (2003).

\bibitem{giovanazzi_pfau}
S.~{Giovanazzi}, A.~{G{\"o}rlitz}, and T.~{Pfau},
\newblock Phys. Rev. Lett. {\bf 89}, 130401 (2002).

\bibitem{zoller_buchler}
H.~P. {B\"uchler {\it et al.}},
\newblock (2006),
\newblock cond-mat/0607294.

\bibitem{cr_dipolar}
J.~{Stuhler} et~al.,
\newblock Phys. Rev. Lett. {\bf 95}, 150406 (2005).

\bibitem{applications_1}
K.~{G{\'o}ral}, L.~{Santos}, and M.~{Lewenstein},
\newblock Phys. Rev. Lett. {\bf 88}, 170406 (2002).

\bibitem{applications_2}
H.~{Pu}, W.~{Zhang}, and P.~{Meystre},
\newblock Phys. Rev. Lett. {\bf 87}, 140405 (2001).

\bibitem{kollath}
A.~Kleine, C.~Kollath, I.~McCulloch, T.~Giamarchi, and U.~Schollwoeck,
\newblock (2007),
\newblock arXiv:0706.0709.

\bibitem{rabl_zoller}
P.~Rabl and P.~Zoller,
\newblock (2007),
\newblock arXiv:0706.3051v1 [quant-ph].

\bibitem{dipolar_theo_earlier_1}
L.~{Santos}, G.~V. {Shlyapnikov}, P.~{Zoller}, and M.~{Lewenstein},
\newblock Phys. Rev. Lett. {\bf 85}, 1791 (2000).

\bibitem{dipolar_theo_earlier_2}
K.~{G{\'o}ral}, K.~{Rza{\.z}ewski}, and T.~{Pfau},
\newblock Phys. Rev. A {\bf 61}, 051601(R) (2000).

\bibitem{dipolar_theo_recent_1}
K.~{Nho} and D.~P. {Landau},
\newblock Phys. Rev. A {\bf 72}, 023615 (2005).

\bibitem{dipolar_theo_recent_2}
S.~{Ronen}, D.~C.~E. {Bortolotti}, D.~{Blume}, and J.~L. {Bohn},
\newblock Phys. Rev. A {\bf 74}, 033611 (2006).

\bibitem{cr_bec}
A.~{Griesmaier}, J.~{Werner}, S.~{Hensler}, J.~{Stuhler}, and T.~{Pfau},
\newblock Phys. Rev. Lett. {\bf 94}, 160401 (2005).

\bibitem{pfau_a0}
T.~Lahaye et~al.,
\newblock Nature (London)  (2007),
\newblock in press.

\bibitem{TG}
M.~Girardeau,
\newblock J. Math. Phys. {\bf 1}, 516 (1960).

\bibitem{Bloch}
B.~{Paredes} et~al.,
\newblock Nature {\bf 429}, 277 (2004).

\bibitem{Weiss}
T.~{Kinoshita}, T.~{Wenger}, and D.~S. {Weiss},
\newblock Science {\bf 305}, 5687 (2004).

\bibitem{giorgini}
G.~E. {Astrakharchik}, J.~{Boronat}, J.~{Casulleras}, and S.~{Giorgini},
\newblock Phys. Rev. Lett. {\bf 95}, 190407 (2005).

\bibitem{dipolar_citro_pra_rc}
R.~Citro, E.~Orignac, S.~De Palo, and M.~L. Chiofalo,
\newblock Phys. Rev. A {\bf 75}, 051602(R) (2007).

\bibitem{giamarchi_general_ref}
T.~Giamarchi,
\newblock {\em Quantum Physics in One Dimension},
\newblock Oxford University Press, Oxford, UK, 2004.

\bibitem{astra}
A.~S. {Arkhipov}, G.~E. {Astrakharchik}, A.~V. {Belikov}, and Y.~E. {Lozovik},
\newblock JETP Lett. {\bf 82}, 39 (2005).

\bibitem{theo_string}
S.~Stringari,
\newblock Phys. Rev. Lett. {\bf 77}, 2360 (1996).

\bibitem{collective_exc_exp_1}
D.~S. Jin, M.~R. Matthews, J.~R. Ensher, C.~E. Wieman, and E.~A. Cornell,
\newblock Phys. Rev. Lett. {\bf 78}, 764 (1997).

\bibitem{collective_exc_exp_2}
D.~M. Stamper-Kurn, H.~J. Miesner, S. Inouye, M.~R. Andrews, and W. Ketterle,
\newblock Phys. Rev. Lett. {\bf 81}, 500 (1998).

\bibitem{RQMC}
S.~Baroni and S.~Moroni,
\newblock Phys. Rev. Lett. {\bf 82}, 4745R (1999).

\bibitem{Stringa_Pita_book}
L.~P. Pitaevskii and S.~Stringari,
\newblock {\em Bose-Einstein Condensation},
\newblock Oxford University Press, USA, 2003.

\bibitem{Chiofalo_sr_charged_bose}
M.~L. Chiofalo, S.~Conti, and M.~P. Tosi,
\newblock J. Phys.: Condens. Matter {\bf 8}, L1921 (1996).

\bibitem{menotti_stringari}
C.~Menotti and S.~Stringari,
\newblock Phys. Rev. A {\bf 66}, 043610 (2002).

\bibitem{haldane_bosons}
F.~D.~M. Haldane,
\newblock Phys. Rev. Lett. {\bf 47}, 1840 (1981).

\bibitem{giamarchi_book_1d}
T.~Giamarchi,
\newblock {\em Quantum Physics in One Dimension},
\newblock Oxford University Press, Oxford, 2004.

\bibitem{kecke05_trapped}
L.~Kecke, H.~Grabert, and W.~Hausler,
\newblock Phys. Rev. Lett. {\bf 94}, 176802 (2005).

\bibitem{ed_long}
E.~Orignac et~al.,
\newblock (2007),
\newblock to be published.

\bibitem{petrov04_bec_review}
D.~Petrov, D.~Gangardt, and G.~Shlyapnikov,
\newblock J. de Phys. IV {\bf 116}, 3 (2004).

\bibitem{sledge}
S.~Pruess and C.~Fulton,
\newblock ACM Trans. Math. Software {\bf 19}, 360 (1993),
\newblock See also the website \texttt{http://www.mines.edu/fs\_home/spruess/}.

\end{thebibliography}

\end{document}